\documentclass[sigconf]{acmart}

\newcommand{\cat}{\ ||\ }

\newcommand{\mve}[1]{\boldsymbol{#1}}

\newcommand{\mset}[1]{\mathcal{#1}}

\newcommand{\mmat}[1]{\boldsymbol{#1}}

\AtBeginDocument{%
  \providecommand\BibTeX{{%
    \normalfont B\kern-0.5em{\scshape i\kern-0.25em b}\kern-0.8em\TeX}}}

\setcopyright{acmcopyright}
\copyrightyear{2022}
\acmYear{2022}
\acmConference[WWW '22] {Proceedings of the ACM Web Conference 2022}{April 25--29, 2022}{Virtual Event, Lyon, France.}
\acmBooktitle{Proceedings of the ACM Web Conference 2022 (WWW '22), April 25--29, 2022, Virtual Event, Lyon, France}
\acmPrice{15.00}
\acmISBN{978-1-4503-9096-5/22/04}
\acmDOI{10.1145/3485447.3511954}

\newcommand{\Set}[1]{\mathcal{#1}}

\newcommand{\Mat}[1]{\mathbf{#1}}

\newcommand{\ie}{\emph{i.e., }}
\newcommand{\eg}{\emph{e.g., }}

\newcommand{\aka}{\emph{aka. }}

\usepackage{bm}
\usepackage{enumitem}
\usepackage{algorithm}
\usepackage{algorithmic}
\usepackage{multirow}
\usepackage{mathtools}
\usepackage{subfigure}
\usepackage{hyperref}
\usepackage{geometry}
\usepackage{graphicx}

\begin{document}


\title{IHGNN: Interactive Hypergraph Neural Network for Personalized Product Search}

\author{Dian Cheng$^{1*}$, Jiawei Chen$^{1*\dagger}$, Wenjun Peng$^1$, Wenqin Ye$^1$, \\Fuyu Lv$^2$, Tao Zhuang$^2$, Xiaoyi Zeng$^2$, Xiangnan He$^{1}$.}

\affiliation{\institution{$^1$University of Science and Technology of China, $^2$Alibaba Group.}} 

\email{{cdboy@mail., cjwustc@, pengwj@mail.}ustc.edu.cn, ywwwwq@pku.edu.cn,}
\email{{fuyu.lfy, zhuangtao.zt, yuanhan}@alibaba-inc.com, xiangnanhe@gmail.com.}

\def\authors{Dian Cheng, Jiawei Chen, Wenjun Peng, Wenqin Ye, Fuyu Lv, Tao Zhuang, Xiaoyi Zeng, Xiangnan He}

\thanks{$*$ Dian Cheng and Jiawei Chen contribute equally to the work.} 
\thanks{$\dagger$ Jiawei Chen is the corresponding author.} 

\renewcommand{\shortauthors}{\authors}

\begin{abstract}
A good personalized product search (PPS) system should not only focus on retrieving relevant products, but also consider user personalized preference. Recent work on PPS mainly adopts the representation learning paradigm, \eg learning representations for each entity (including user, product and query) from historical user behaviors (\aka user-product-query interactions). However, we argue that existing methods do not sufficiently exploit the crucial \textit{collaborative signal}, which is latent in historical interactions to reveal the affinity between the entities. Collaborative signal is quite helpful for generating high-quality representation, exploiting which would benefit the representation learning of one node from its connected nodes.

To tackle this limitation, in this work, we propose a new model IHGNN for personalized product search. IHGNN resorts to a hypergraph constructed from the historical user-product-query interactions, which could completely preserve ternary relations and express collaborative signal based on the topological structure. On this basis, we develop a specific \textit{interactive hypergraph neural network} to explicitly encode the structure information (\ie collaborative signal) into the embedding process. It collects the information from the hypergraph neighbors and explicitly models neighbor feature interaction to enhance the representation of the target entity. Extensive experiments on three real-world datasets validate the superiority of our proposal over the state-of-the-arts.


\end{abstract}



\maketitle

\section{Introduction}


Online shopping pervades our daily lives. As the number of products in e-shopping platforms grows explosively, it is almost impossible for a user to discover desirable products without the help of product search engines. The search engine would retrieve a list of potential products for each user when he submits a query. The quality of the search results is crucial for both user satisfaction and retailer revenues.

Different from the traditional search task that focuses on finding the items matching the query, product search is more challenging as the target products are highly personalized~\cite{ge2018personalizing,teevan2008personalize}. In a typical e-shopping scenario, it is common that users have quite different purchase intents even if they issue the same query. Taking the query ``Delicious Food'' as an example, European users may expect some Pasta while Chinese users may be interested in dumplings. It is widely recognized that user purchases would be affected by their personalized preference \cite{wu2022survey,hem}. Therefore it is important for a product search engine to be personalized, with the goal to ``understand exactly what the user wants and give him personalized suggestions''~\cite{guo2019attentive}.

To achieve this goal, existing methods on personalized product search (PPS)~\cite{hem,zam,tem,ge2018personalizing,lu2019psgan,yao2020rlper,zhou2020enhancing,zhou2020encoding} mainly adopt the representation learning paradigm. They transform each entity (including user, product, and query) to a vectorized representation and then predict user purchase inclination based on the embeddings. Despite their decent performance, we argue that existing methods have not sufficiently exploited the \textit{collaborative signal}. 
It is latent in historical user-product-query interactions to reveal the affinity among the entities, which is crucial for personalized search. For example, the users engaging with the same product may have similar preferences; the queries under which the same products are purchased by a user may have similar semantics. 
When equipped with such rich affinity information, the representation learning of one node can benefit from other related ones, resulting in higher-quality representations. 
The work that is closest to ours is \cite{srrl}, however, it only exploits three manually-designed affinity patterns, which is far from sufficient. How to fully leverage the collaborative signal for PPS is still an open problem.
 
\begin{figure}[t!]
	\centering
	\vspace{-0.35cm}
    \subfigure[Bipartite graph]{
        \label{graph_gcn}
		\raisebox{0.2\height}{
			\includegraphics[height=0.27\linewidth]{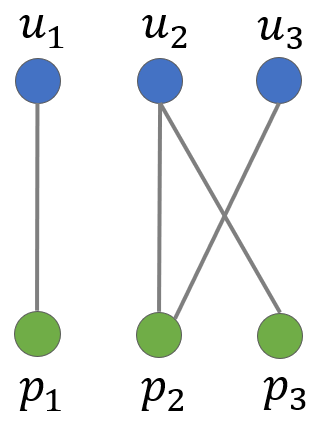}
		}
    }
	\subfigure[Collapsed Graph]{
		\label{graph_cmp_1}
		\raisebox{0.2\height}{
			\includegraphics[height=0.27\linewidth]{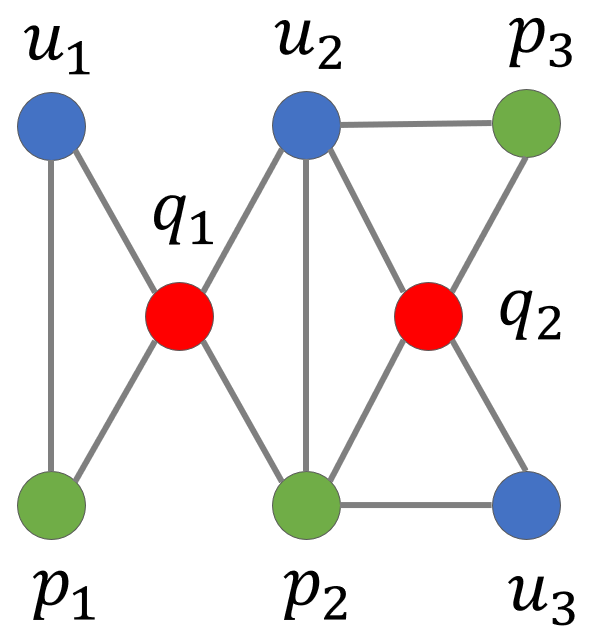}
		}
    }
	\subfigure[Hypergraph]{
		\label{graph_cmp_2}
		\includegraphics[height=0.4\linewidth]{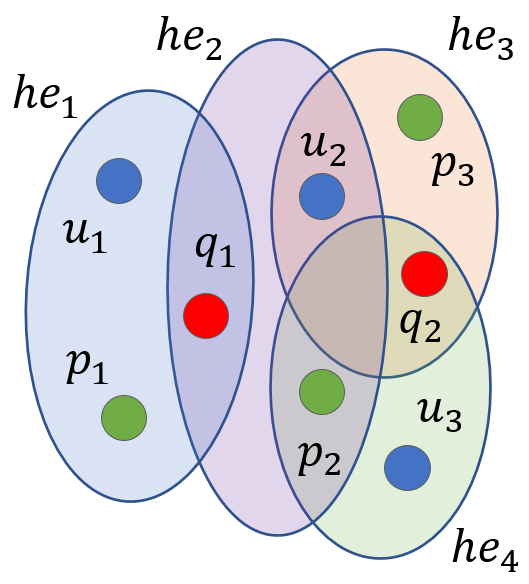}
    }
    \setlength{\abovecaptionskip}{-0cm}
  \setlength{\belowcaptionskip}{-0.1cm}
    \caption{
    (a) gives a sample of user-product graph in recommendation, while (b)(c) give samples of collapsed graph and hypergraph in personalized product search. $u_i, q_i, p_i (i = 1, 2,3)$ are nodes representing user, product or query respectively, $he_i (i = 1, 2, 3,4)$ denotes a hyperedge shown as an ellipse in (c). Each user-query-product interaction corresponds to a hyperedge. There are four interactions: $(u_1, q_1, p_1)$, $(u_2, q_1, p_2)$, $(u_2,q_2,p_3)$ and $(u_3, q_2, p_2)$.
	}
	\label{graph_cmp}
\end{figure}

This work fills the research gap.
Being aware of the effectiveness of graph neural network (GNN) for relational representation learning~\cite{graphsage,gcn}, we wish to take its advantages for PPS. In traditional search~\cite{lse,zhang2018multiresolution} and recommendation~\cite{lightgcn, SGL}, the graph is bipartite that presents query-word matching and user-item matching. Performing graph convolution on these graphs can collect information from similar neighbors, which strengthens the representations with collaborative signal explicitly. However, the appealing idea is non-trivial to transfer to the PPS task, due to two main difficulties:

(P1) More complicated than traditional search and recommendation, interactions in PPS are ternary rather than binary --- each interaction involves three elements: user, product and query. It is intractable to construct a simple graph to preserve ternary relations. For example, if we forcibly split the ternary relations into three binary relations among users, products and queries \cite{he2015trirank}, we will lose the information like under which query the user-product interaction happens. Figure \ref{graph_cmp_1} gives a toy example, we cannot determine under which query the interaction $(u_2, p_2)$ happened: $q_1$, or $q_2$, or both of them. Since a simple graph cannot losslessly represent the ternary relations, we need to resort to a more general topological structure for model development.

(P2) Existing GNNs mainly adopt a linear aggregation over the features of neighbors, ignoring the high-order feature interactions of neighbors. In fact, in PPS, the interactions between related entities could be a strong signal to indicate the characteristics of the target node.
For example, when a user searches for ``women's bag'' and finally purchases a bag of the brand ``Hermès'', the interaction of the query and product would generate a quite useful semantic (\eg ``women's luxury brands'') for profiling the user's preference. We need to explicitly consider feature interaction to enhance the representation for PPS.

To tackle these problems, we propose to construct a \textit{hypergraph} from the ternary user-product-query interactions. Compared with simple graph, hypergraph is a more suitable data structure for modeling ternary relations, because each hyperedge can connect any quantity of nodes. 
On this basis, we further propose a novel PPS model, named Interactive HyperGraph Neural Network (IHGNN), which recursively aggregates neighbor information along the aforementioned hypergraph. Distinct to the GNNs for recommendation or non-personalized search, our IHGNN makes two important improvements: (1) As each hyperedge in the hypergraph connects multiple nodes, IHGNN adopts a two-step information propagation scheme --- node aggregation, which aggregates the information from the connected nodes to update the hyperedge representation; and hyperedge aggregation, which collects the information from the related hyperedges to update the target node representation. (2) As the neighbor interaction is important in PPS, we explicitly conduct high-order feature interaction of neighbors, and then aggregate the interacted results to enhance the target node representations.


In summary, this work makes the following contributions:

\begin{itemize}
\item We approach the PPS task with a user-product-query hypergraph and develop a hypergraph neural network to explicitly encode the collaborative signal into representation learning.
\item We highlight the importance of exploiting feature interaction in representation learning, and propose to explicitly model high-order feature interactions of neighbors in hypergraph embedding aggregation.
\item We conduct extensive experiments on three real-world datasets to demonstrate the effectiveness and the rationality of each component design of IHGNN.
\end{itemize} 

\section{Preliminary}
In this section, we first give the formal definition of personalized product search task, and then give some background knowledge related to hypergraph.
\subsection{Personalized Product Search}

Suppose we have a product search engine with a user set $\Set U$, an item set $\Set P$, a possible query set $\Set Q$ and a set of historical user-product-query interactions $\Set L$. Let $u$ ($p$, or $q$) denote a user (an item, or a query) in $\Set U$ ($\Set P$, or $\Set Q$). $\Set L$ consists of a list of user-item-query triples $(u,p,q)$, indicating user $u$ has purchased\footnote{We remark that here the ``purchase'' can be replaced by other type of implicit feedback such as ``
Click'' or ``Add-to-Cart''. In this work we simply use the word ``purchase'' as a placehoder for better description.} the product $p$ under the query $q$. Also, we use $y_{upq}\in \{0,1\}$ to indicate whether the interaction $(u,p,q)$ happens, \ie $y_{upq}=1$ for $(u,p,q)\in \Set L$ and $y_{upq}=0$ for $(u,p,q) \notin L$. The goal of PPS is to learn a score function $f: \Set U \times \Set P \times \Set Q \to R$ to accurately predict the probability of a user $u$ to purchase product $p$ when searching query $q$. 

\subsection{Hypergraph}

Different from simple graph, a hypergraph is a more general topological structure where a hyperedge could connect two or more nodes. Let $\Set G=(\Set V, \Set E, \Mat H)$ be an instance of hypergraph, which includes a node set $\Set V$ and a hyperedge set $\Set E$. The $|\Set V|\times |\Set E|$ incidence matrix $\Mat H$ describes the connectivity of the hypergraph, with entries defined as:
\begin{equation}
    \begin{split}
        h(v, e) = \left\{
            \begin{aligned}
                \ &1 , \quad {\rm if} \  e \text{ connects } v, \\
                \ &0 , \quad {\rm if} \  e \text{ disconnects } v,
                \end{aligned}
        \right.
    \end{split}
\end{equation}
On this basis, we further give some notations in a hypergraph. For each node $v \in \Set V$, its degree is defined as $d(v) = \sum\limits_{e \in \Set E} {h(v,e)}$; For each edge $e \in \Set E$, its degree is $d(e) = \sum\limits_{e \in \Set E} {h(v,e)}$. They can be further collected as two diagonal matrices $D_v$ and $D_e$ of node degrees and edge degrees, respectively. Let $\Set E_v$ denote a set of related hyperedges that connect to the node $v$ (\ie $\Set E_v=\{e\in \Set E|h(v,e)=1\}$), $\Set V_e$ denote a set of nodes to which the hyperedge $e$ connects(\ie $\Set V_e=\{v\in \Set V |h(v,e)=1\}$). Also, we can define the ``neighbors'' ($N_v$) of node $v$ as a set of nodes that share at least one hyperedge with the node $v$ (\ie $N_v=\{a \in \Set V|\exists e \in \Set E, h(v,e) = 1\& h(a,e) = 1\}$).

\begin{figure}[t!]
    \centering
    \includegraphics[width=0.28\textwidth]{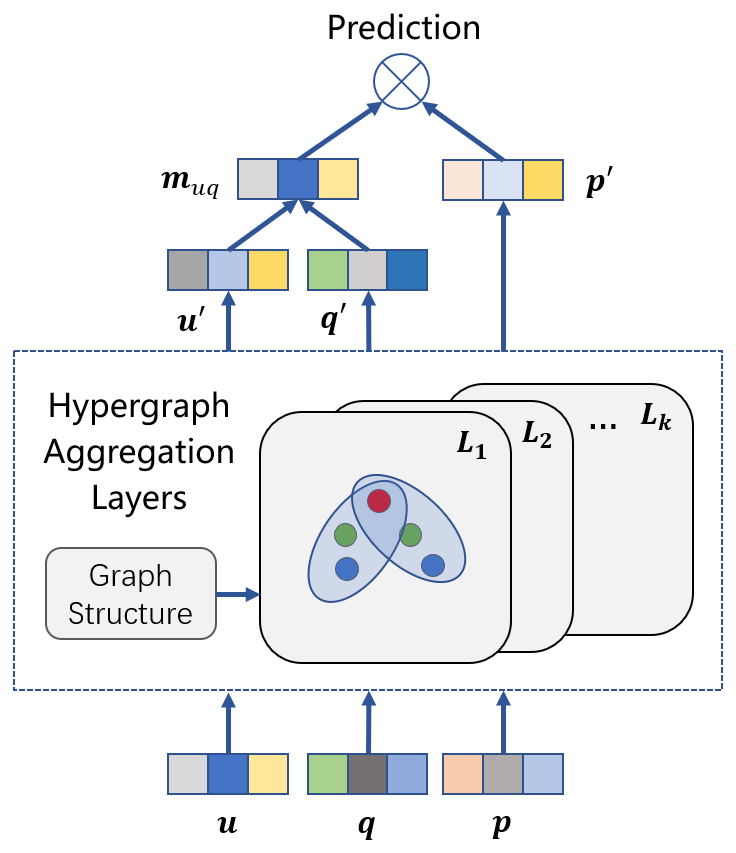}
    \setlength{\abovecaptionskip}{-0cm}
  \setlength{\belowcaptionskip}{-0.2cm}
    \caption{
        Illustration of the proposed IHGNN. 
    }
    \label{model_arch}
\end{figure}
 
 
\begin{figure*}
	\centering
    \subfigure[vanilla hypergraph propagation]{
        \label{hgcn}
        \raisebox{0.17\height}{\includegraphics[width=0.47\linewidth]{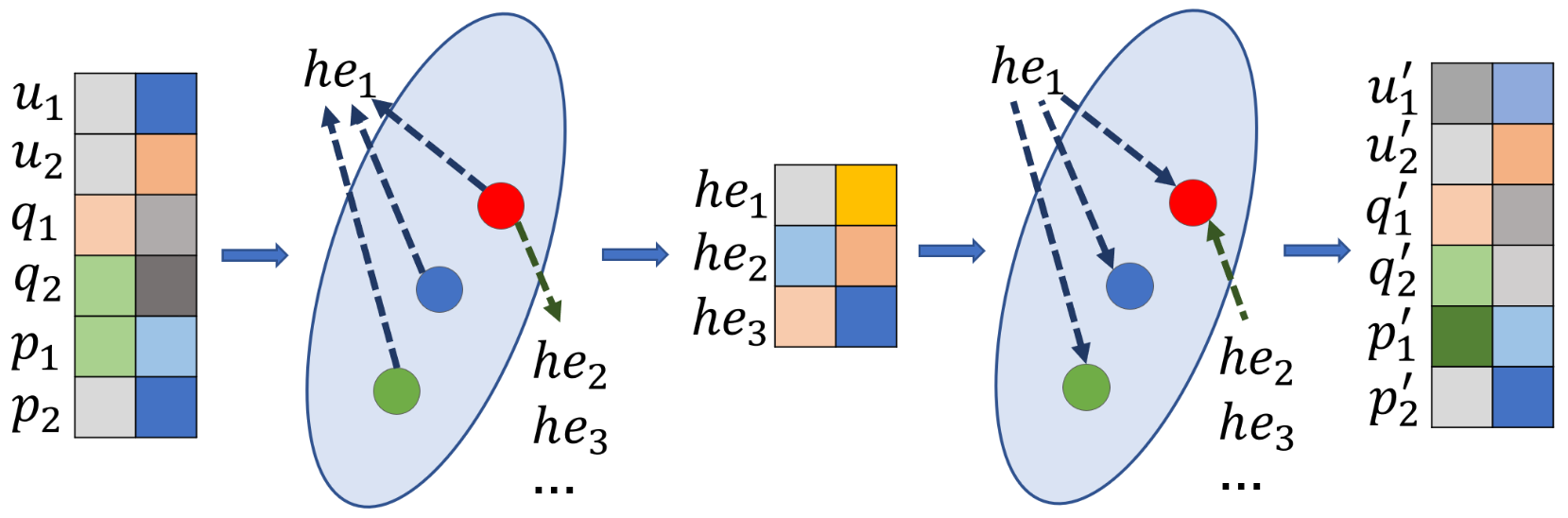}}
    }
    \hspace{1cm}
    \subfigure[Improved hypergraph propagation with higher-order interaction]{
		\label{ihgnn}
		\includegraphics[width=0.43\linewidth]{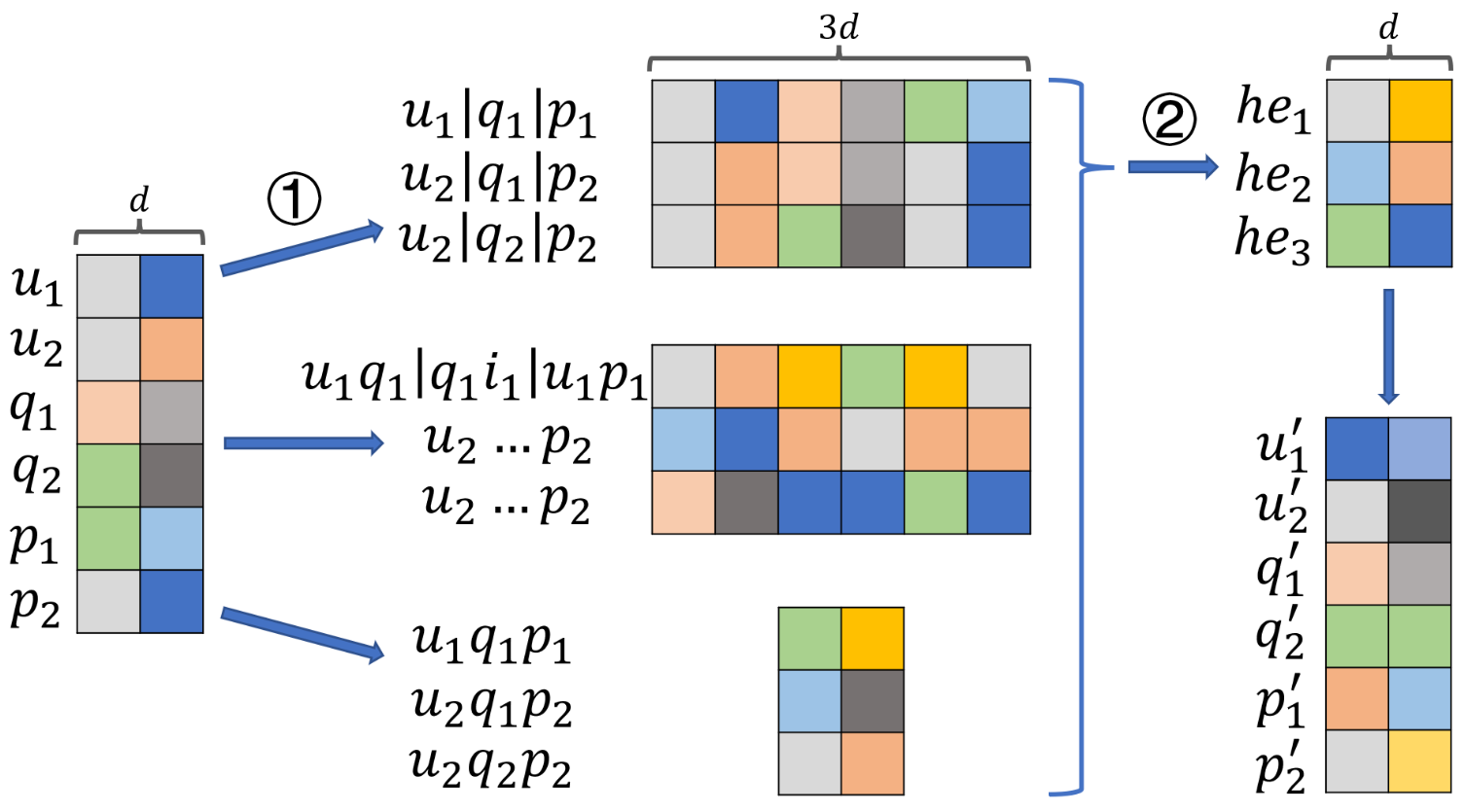}
    }
    \setlength{\abovecaptionskip}{-0cm}
  \setlength{\belowcaptionskip}{-0.1cm}
    \caption{
        Illustration of how IHGNN aggregates the information from neighbors.
    }
	\label{cmp_ihgnn_hgcn}
\end{figure*}
\section{METHODOLOGY}
In this section, we detail the proposed IHGNN for personalized product search. IHGNN aims at explicitly encoding collaborative signal into representation using user-product-query hypergraph. IHGNN contains four modules: (1) a hypergraph construction module that constructs hypergraph from the historical user-product-query interactions $\Set L$; (2) an embedding generation module that transforms the features of entities into their initial representations; (3) an aggregation module that refines the embeddings by collecting the information from neighbors and high-order neighbors; and (4) a prediction module that generates the prediction of user purchase inclination based on the refined embeddings. Finally, we discuss the properties of IHGNN and its connection with existing methods.

\subsection{Hypergraph Construction}
\label{sec_hypergraph_construction}
We first construct a hypergraph $\Set G$ based on historical ternary user-product-query interactions. Specifically, given historical interactions $\Set L$, we have a hypergraph $\mset{G} = (\mset{V}, \mset{E}, \mmat{H})$ including: (1) a node set $\Set V$ whose element represents a user, product or query (\ie  $\mset{V} = \mset{U} \cup \mset{Q} \cup \mset{P}$); (2) a hyperedge set $\Set E$, whose element represents a ternary relation depicting there exists an historical interaction among them (\ie $e \in \Set E \leftrightarrow (u,p,q) \in \Set L$, $h(u,e) = 1,h(p,e) = 1,h(q,e) = 1$).

The constructed hypergraph could preserve complete ternary relations and capture collaborative signal via topological structure. From which, we can easily deduce nodes' affinity based on their proximity in the hypergraph. 
Figure \ref{graph_cmp_2} gives a toy example. We can deduce user $u_1$ and $u_2$ may have similar preferences as they both connect to (issued) the query $q_2$.
Such hypergraph provides an opportunity to fully exploit collaborative signal for PPS, \eg we can leverage hypergraph embedding to encode such important signal (\ie hypergraph structure) into representation. 
We next introduce how to utilize the hypergraph structure for enhancing PPS.

\subsection{Embedding Generation Module}

\newcommand{\localwq}{\mve{w}_q}
\newcommand{\localwqinq}{\localwq \in q}
\newcommand{\localqset}{\{\localwq | \localwqinq\}}
\newcommand{\localqaverage}{\frac{\sum_{\localwqinq} \  \localwq}{|q|}}

This module aims at mapping entities into a common representation space. Here we follow previous work \cite{vulic2015monolingual} to transform the features of queries to their representations. That is, for each query $q\in \Set Q$, we utilize the query content information and use a mean pooling over word embeddings $z_w$ to generate the query embedding $z_q\in R^d$:
\begin{equation}
    \begin{split}
        z_q = \frac{\sum_{w\in q} \  z_w}{|q|}
    \end{split}
\end{equation}
For each user $u \in \mset{U}$ (or product $p \in \mset{P}$), we directly generate its embedding $z_u\in R^d$ (or $z_p\in R^d$) from an embedding look-up table $\Mat E\in R^{((|\Set U|+|\Set P|)\times d)}$ (\ie a parameter matrix).

\subsection{Aggregation Module}
In this module, we describe our embedding aggregation scheme, which iteratively aggregates the information from the neighbors to encode the collaborative signal into the embeddings. We first illustrate the design of one-layer aggregation and then generalize it to multiple successive layers.

\subsubsection{First-order Aggregation}
Intuitively, hypergraph neighbors provide useful signals to understand the target node's characteristics. For example, the interacted products or submitted queries provide direct evidences on a user's preferences; Analogously, relevant products are quite helpful to depict the semantics of the query. Thus, it is nature to collect the information from the neighbors to refine the representation of the target node. Distinct to the GNNs for recommendation \cite{wang2019neural} or non-personalized search \cite{zhang2019neural}, our IHGNN is conducted on hypergraph and thus adopts a two-step information propagation scheme --- \ie hyperedges serve as mediums to process and transfer the information from the neighbors. To be more specific, for each node $v$, as shown in Figure \ref{hgcn}, we perform the following two operations to update its representation:

\textbf{Node aggregation.}
In this stage, we aggregate the information propagated from $v$'s neighbors to related hyperedge. Specifically, for each $e\in \Set E_v$, we define the aggregation function as follow:
\begin{equation}
    \begin{split}
{m_e} = [{z_u}\cat {z_p}\cat {z_q}]\Mat W
    \end{split}
    \label{eq_ihgnn_p1_o1_raw}
\end{equation}
where $m_e$ denotes the information contained by the medium $e$, $u,p,q$ denote the user, product, query node that $e$ connects, $||$ denotes the concatenation operation, $\Mat W \in \mathbb R^{3d \times d}$ is a trainable weight matrix to distill useful knowledge from the neighbor nodes. Here we choose linear transformation instead of the conventional simple average, as the hyperedge would collect different types of nodes and we believe they may make different contributions.

\textbf{Hyperedge aggregation.}
In this stage, we aggregate the information from the related hyperedge to refine the representation of node $v$ as:
\begin{equation}
    \begin{split}
        z_v^{(1)} = \frac{ \sum_{e \in \Set E_v} {m_e } }{d(v)}
    \end{split}
\end{equation}
With such two operations, the knowledge of neighbor nodes has been explicitly injected into the target node's representation, reinforcing its quality.

\textbf{Modeling neighbor feature interaction.}
However, we argue that such linear aggregation scheme is insufficient, ignoring the knowledge from the feature interaction among the neighbors. In fact, in PPS, the neighbor interaction could be a strong signal that indicates the characteristics of the target node. For example, provided a user searches for ``women's bag'' and finally purchases a bag of the brand ``Hermès'', the interaction between the query and the product would generate a quite useful semantic (\eg ``women's luxury brands'') for profiling the user's preference. Based on this point, we further develop an improved node aggregation operation as shown in Figure \ref{ihgnn}, which explicitly models the feature interaction among the nodes (\ie $u,p,q \in \Set V_e)$ that the hyperedge $e$ connects. We have:
\begin{equation}
    \begin{split}
        f^{o1}_e &= z_u \cat z_q \cat z_p \\
        f^{o2}_e &= (z_u \odot z_q) \cat (z_u \odot z_p) \cat (z_q \odot z_p) \\
        f^{o3}_e &= z_u \odot z_q \odot z_p
    \end{split}
\end{equation}
Where $\odot$ stands for element-wise product, $f^{o1}_e,f^{o2}_e,f^{o3}_e$ respectively capture 1,2,3-order feature interaction of neighbors. Analogously, we use a linear layer to aggregate the interaction information as:
\begin{equation}
    \begin{split}
        m_e = [f^{o1}_e \cat f^{o2}_e \cat f^{o3}_e] \ \mmat{W}_{o3}
    \end{split}
    \label{eq_ho123}
\end{equation}
This useful information will be further propagated into the target node representation via the hyperedge aggregation. We remark that here we just conduct feature interaction among the neighbors that share common hyperedge (\ie $u,p,q \in \Set V_e \& e\in \Set E_v$) rather than among all neighbors (\ie $u,p,q \in \Set N_v$). The reason is that these specific neighbors are highly related and their interactions potentially generate strong signal. Also, this treatment is much more efficient. Conducting feature interaction among all neighbors is usually computationally unavailable.

\subsubsection{Higher-order Aggregation}
To completely exploit the collaborative signal, we further consider to stack aforementioned aggregation module such that the target node's representation could benefit from the high-order neighbors. In fact, although these nodes are not directly connected with the target node, they indeed share some similarity and provide useful knowledge to learn the target representation. Concretely, As Figure \ref{ihgnn} displays, in the $l$-th layer, the representation of the node $v$ is recursively updated as follow:
\begin{equation}
    \begin{split}
        z_v^{(l)} = \frac{ \sum_{e \in \Set E_v} {m_e^{(l)} } }{d(v)}
    \end{split}
\end{equation}
where the hyperedge information $m_e^{(l)}$ can be calculated from $z_u^{(l-1)}$, $z_p^{(l-1)}$,$z_q^{(l-1)}$ with equation (\ref{eq_ihgnn_p1_o1_raw}) or (\ref{eq_ho123}). Finally we concatenate the representations learned by different layers to generate the entity's final representation:
\begin{equation}
    \begin{split}
z_v^{*}=z_v^{(0)}||z_v^{(1)}||...||z_v^{(L)}
    \end{split}
\end{equation}
By doing so, we enrich the initial embeddings with the information propagated from similar (high-order) neighbors. The collaborative signal has been explicitly injected in the representation.

\subsection{Prediction Module}

\newcommand{\localmuq}{\mve{M}_{\mve{uq}}}
This model aims at generating prediction based on the learned embeddings. As this module is not our focus, we simply refer to \cite{hem} for implementation. It is relatively simple but proved effective. Specifically, we estimate the purchase probability of user $u$ towards the product $p$ when searching query $q$ as follow:
\begin{equation}
    \begin{split}
{{\hat y}_{uqp}} = \text{sigmoid}\left({(\lambda z_u^* + (1 - \lambda )z_q^*)^T}z_p^*\right)
    \end{split}
\end{equation}
where $\lambda$ is a hyper-parameter controlling the contribution from the user preference and item relevance.

\subsection{Model Optimization}
For fairly comparison, we closely follow the related work \cite{srrl} to learn the model parameters. Specifically, we optimize the following objective function:
\begin{equation}
    \begin{split}
        L = \sum_{(u, q, p) \in \Set L \cup \Set B^-} -\, y_{uqp} \cdot \log \hat{y}_{uqp} - (1 - y_{uqp}) \cdot \log (1 - \hat{y}_{uqp})
    \end{split} \nonumber
\end{equation}
where $\Set B^-$ denotes a negative sampled set that contains the triples with $y_{uqp}=0$.

\subsection{Discussion}

\subsubsection{Connection with existing PPS methods.}
First, we compare our IHGNN with GraphSRRL \cite{srrl}. GraphSRRL explicitly considers three manually-designed affinity patterns --- \eg Provided there are two interactions $(u_1,q,p)$, $(u_2,q,p)$, $u_1$ and $u_2$ may have similar preference. In fact, these affinity patterns can adaptively be captured by our model. With embedding aggregation, the information of $z_q$, $z_p$ as well as their interactions $z_q \cdot z_p$ can be propagated into the representation of $u_1$ and $u_2$, making $u_1$ and $u_2$ hold a certain similarity. Similar analysis can be conducted for other patterns. Besides such three patterns, our IHGNN captures more and thus yields better performance than GraphSRRL as reported in our experiments.

Also, it is worth to discuss the connections with some related work \cite{guo2019attentive,bi2019leverage} that models Long Short-Term Preference. These methods would deduce user long-term (or short-term) preference from his purchased products (or submitted queries). In fact, it can be considered as a special way of utilizing collaborative signal. Our IHGNN can also capture this pattern in the one-layer aggregation. We would collect the information from the neighbors to update the representation.

\subsubsection{Connection with existing hypergraph neural network.}
Recent years also witnessed some work \cite{hgcn,hgcn2} on hypergraph neural network. These methods derive HGNN from spectral convolution and it has similar propagation scheme as our IHGNN. But IHGNN differs from HGNN in the following two aspects: (1) when performing node aggregation, we give different weights for different types of nodes; (2) we explicitly model feature interaction of neighbors, which is of importance in PPS. 

\subsubsection{Connection with graph neural network.}
Here we mainly compare our IHGNN with the GNNs on the collapsed graph, where we split the ternary relations into three binary
relations. The key difference is in that our IHGNN utilize a medium  (\ie hyperedge) to process and transfer information, such that the ternary relation can be treated in a holistic view; while GNN can only handle the fragment information. It makes our IHGNN usually achieve better performance than GNNs on the collapsed graph.



\section{Experiments}

In this section, we conduct experiments to validate the effectiveness of our IHGNN. Our experiments are intended
to address the following research questions:
\begin{itemize}
\item \textbf{RQ1}: How does IHGNN perform compared with state-of-the-art PPS methods?
\item \textbf{RQ2}: How do different components (\eg using hypergraph, weighted propagation, high-order interaction) contribute to the model performance?
\item \textbf{RQ3}: How do different hyper-parameters (\eg depth of aggregation $L$, and embedding size $d$) affect the performance of IHGNN?
\end{itemize}


\subsection{Datasets}
We utilize three datasets in our experiments, including one real-world dataset that collected from real PPS scenario and two available conventional semi-synthetic dataset.

\textbf{AlibabaAir dataset}. 
The dataset is collected from taobao.com, one of the biggest e-shopping platforms in China. 
We randomly sampled 200,000 users and collect their search logs on the platform from 23-Dec-2020 to 29-Dec-2020. 
The dataset contains the display information of user-query-product, and labels denoting whether the user clicked the product.
Approximately 930,000 clicks and 10,000 purchases are contained.
This dataset also contains basic query word segmentation results.
Users, query words/characters, products were all encrypted as id numbers for privacy protection. We conduct 5-core and 1-core filtering (\ie retaining entities with at least 5 or 1 interactions) to generate two datasets, marked as Ali-1Core and Ali-5Core, respectively. This treatment would like to show the robust of our model on cold start users (or items).

\textbf{CIKMCup dataset}. 
This is a product search dataset from CIKMCup2016 Track2. 
However, over half of the search logs are anonymous and do not have user ids. 
Another shortcoming is that most queries are manually composed according to category browsing. 
In our experiments, in order to use enough data, we preserve the category browsing results, click data, and view data after we filter out the anonymous search logs. Also, 5-core filtering has been conducted.
Note that GraphSRRL~\cite{srrl} directly used the default recommendation results as ground truth, which may not be consistent with user true preference and thus is not adopted by us.

\textbf{Amazon simulated dataset}. This is a review dataset consisting of users' reviews on products. It is used by \cite{hem} in earlier research of PPS task. 
A user-product review pair is extended to a user-query-product triple by treating the product's category string as a query. 
Since Amazon datasets are semi-synthetic and are not as convincing as other datasets (AlibabaAir) adopted in this work, here we simply choose one typical sub-dataset (\ie CDs\_5) for experiments.


The statistics of three datasets are shown in Table \ref{ds_info}.

\begin{table}[!tb]
    \centering
    \caption{Dataset Statistics.}
    \begin{tabular}{cccccc}
      \toprule
            & users & queries & products & interactions \\
      \midrule
      Ali-1Core & 200,000 & 102,816 & 220,779 & 940,946 \\
      \midrule
      Ali-5Core & 39,976 & 46,098 & 39,326 & 403,982 \\
      \midrule
      CIKM & 23,882 & 3,256 & 23,550 & 339,341 \\
      \midrule
      CDs\_5 & 112,379 & 509   & 67,602 & 1,296,885 \\
      \bottomrule
      \end{tabular}
    \label{ds_info}
\end{table}

\begin{table*}[!tb]
  \centering
  \caption{
    Performance comparisons for personalized product search. The boldface font denotes the winner in that column.
    Also, the best baselines are marked with $\dagger$. The row `Gain' indicates the relative performance gain of our IHGNN compared to the best baseline. `*' and `**' denote the improvement is significant with t-test with $p < 0.05$ and $p < 0.1$, respectively.
  }
  \resizebox{\textwidth}{!}{
    \begin{tabular}{c|ccc|ccc|ccc|ccc}
      \toprule
            & \multicolumn{3}{c|}{Ali-1Core} & \multicolumn{3}{c|}{Ali-5Core} & \multicolumn{3}{c|}{CIKM} & \multicolumn{3}{c}{CDs\_5} \\
            & HR@10 & NDCG@10 & MAP@10 & HR@10 & NDCG@10 & MAP@10 & HR@10 & NDCG@10 & MAP@10 & HR@10 & NDCG@10 & MAP@10 \\
      \midrule
      LSE   & 0.1373  & 0.0848  & 0.0870  & 0.1979  & 0.1257  & 0.1343  & 0.3147  & 0.2215  & 0.2735  & 0.1300  & 0.0696  & 0.0515  \\
      HEM   & 0.1490  & 0.0957  & 0.0980  & 0.2238  & 0.1485  & 0.1610  & 0.4100  & 0.3541  & 0.4601  & 0.1409  & 0.0762  & 0.0567  \\
      ZAM   & 0.1412  & 0.0915  & 0.0978  & 0.2559  & 0.1749  & 0.1916  & 0.3767  & 0.3273  & 0.4491  & 0.0938  & 0.0482  & 0.0345  \\
      TEM   & 0.1320  & 0.0916  & 0.1020  & 0.2493  & 0.1758  & 0.1939  & 0.4471  & 0.3576  & 0.4509  & 0.1292  & 0.0666  & 0.0478  \\
      GraphSRRL & 0.1306  & 0.0872  & 0.0952  & 0.2445  & 0.1659  & 0.1811  & 0.4729  & 0.3729  & 0.4700  & 0.1577  & 0.0846  & 0.0625  \\
      \midrule
      GCN   & 0.1773  & 0.1177  & 0.1221  & 0.2549  & 0.1743  & 0.1876  & 0.4424  & 0.3557  & 0.4394  & 0.1451  & 0.0785  & 0.0584  \\
      GAT   & 0.1908$^\dagger$ & 0.1264$^\dagger$ & 0.1329$^\dagger$ & 0.2715$^\dagger$ & 0.1858$^\dagger$ & 0.2018$^\dagger$ & 0.5238$^\dagger$ & 0.4211$^\dagger$ & 0.5217$^\dagger$ & 0.1652$^\dagger$ & 0.0873$^\dagger$ & 0.0638$^\dagger$ \\
      HyperGCN & 0.1803  & 0.1180  & 0.1212  & 0.2655  & 0.1826  & 0.1986  & 0.4100  & 0.3474  & 0.4430  & 0.1396  & 0.0755  & 0.0561  \\
      \midrule
      IHGNN-O3 & \textbf{0.2073}* & \textbf{0.1465}* & \textbf{0.1591}* & \textbf{0.2894}* & \textbf{0.2091}* & \textbf{0.2307}* & \textbf{0.5514}* & \textbf{0.4855}* & \textbf{0.6314}* & \textbf{0.1672}* & \textbf{0.0893}** & \textbf{0.0657}** \\
      \midrule
      Gain  & 8.6\% & 15.9\% & 19.7\% & 6.6\% & 12.5\% & 14.3\% & 5.3\% & 15.3\% & 21.0\% & 1.2\% & 2.3\% & 3.0\% \\
      \bottomrule
      \end{tabular}}
  \label{ret_overall}%
\end{table*}%

\subsection{Compared Methods}

The following methods are tested in our experiments:
\begin{itemize}
\item  \textbf{LSE} \cite{lse} is a classic latent vector space model for product search. LSE learns word and item representations in one latent vector space and directly models the match score between products and query words. It does not consider to model users and therefore is non-personalized.
\item \textbf{HEM} \cite{hem} extends LSE \cite{lse} with personalization setting. HEM adopts representation learning paradigm. It learns user, product and query representation and then utilize an inner product function to make a prediction. 
\item \textbf{ZAM} \cite{zam} extends HEM to determine how much personalization is needed. 
\item \textbf{TEM} \cite{tem} is a transformer-based embedding model for personalized product search. It improves ZAM by adding the ability of dynamically controlling the influence of personalization. Its core idea is to encode the sequence of query and user's purchase history with a transformer architecture, which gives different personalization weights to different history items.
\item \textbf{GraphSRRL} \cite{srrl} improves HEM with utilizing three manual-designed affinity patterns to enhance the representation learning.
Because GraphSRRL demonstrate superiority over DREM \cite{drem}, we do not include DREM as baselines.
\item \textbf{GCN \cite{gcn}, GAT \cite{gat}, HyperGCN \cite{hgcn}}: We also design three quite competitive baselines --- utilizing graph-based methods for enhancing representation learning. GNN and GAT are conducted on collapsed graph, while HyperGCN is conducted on the same hypergraph as ours. We remark that here we do not compare some other graph-based  methods \cite{niu2020dual}, as they require other side information, which is unfair. 
\item \textbf{IHGNN}: the proposed method in this paper. We test different version of IHGNN, where `IHGNN-O*' denotes the model considering different maximum orders of feature interaction. 
\end{itemize}

\subsection{Experimental Setup}

\subsubsection{Data Split.} 
We split the dataset into 3 parts according to the search time of user interaction: 70\% for training, 10\% for validation, and 20\% for testing.

\subsubsection{Evaluation} 
For each test data (user, query, products), we use the top-10 setting, where the top-10 ranking results are utilized to calculate the evaluation metrics. 
We use three conventional metrics: hit ratio (HR), normalized discounted cumulative gain (NDCG) and mean average precision (MAP). 
HR focuses on the retrieval performance by calculating the ratio of positive products appearing in search results. 
NDCG and MAP are common metrics for ranking algorithms.

\subsubsection{Implementation Details} 
We implement our IHGNN~\footnote{https://github.com/CDboyOne/IHGNN} with PyTorch.
The embedding size is set to 32.
We randomly sample 10 negative products for one user-query-product interaction.
We use Adam optimizer with learning rate 0.001 to optimize all models, where the batch size is 100. 
The embeddings are initialized by using the default Xavier uniform method and other parameters are initialized by the Kaiming uniform method. 
All graph-based models, including GraphSRRL, GCN, GAT, HyperGCN, and IHGNN, use 2 graph layers.
Models are trained on i9-9900X CPU and 2080Ti GPU. 
In training process, we calculate the metrics on validation and test dataset.
We choose the test metrics according to the best NDCG@10 result on validation set. 
We train all models until they converge for fair comparison.
Hyper-parameter $\lambda$ is set to $0.5$.

\begin{figure*}
	\centering
  \subfigure[Ali-1Core NDCG@10]{
    \label{fig_ablation_gnns_1C_NDCG}
    \includegraphics[width=0.235\linewidth]{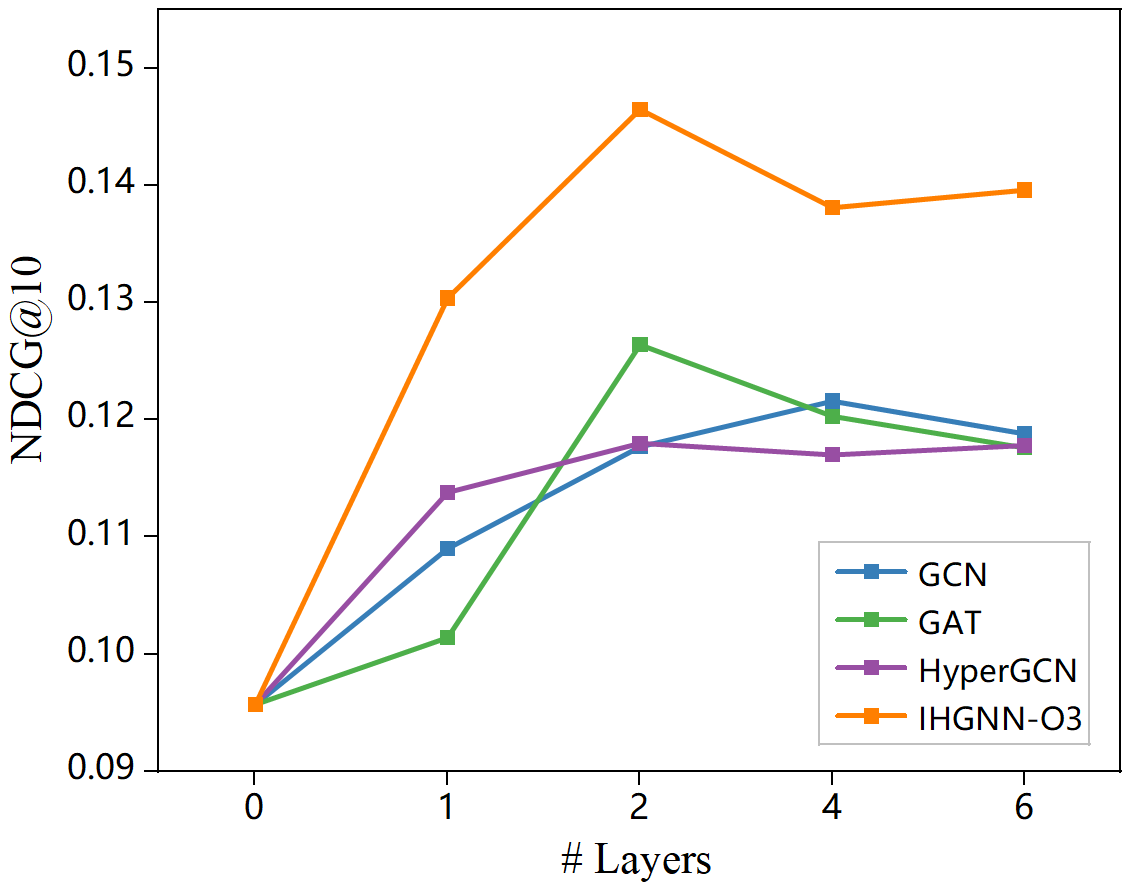}
  }
  \subfigure[Ali-5Core NDCG@10]{
    \label{fig_ablation_gnns_5C_NDCG}
    \includegraphics[width=0.235\linewidth]{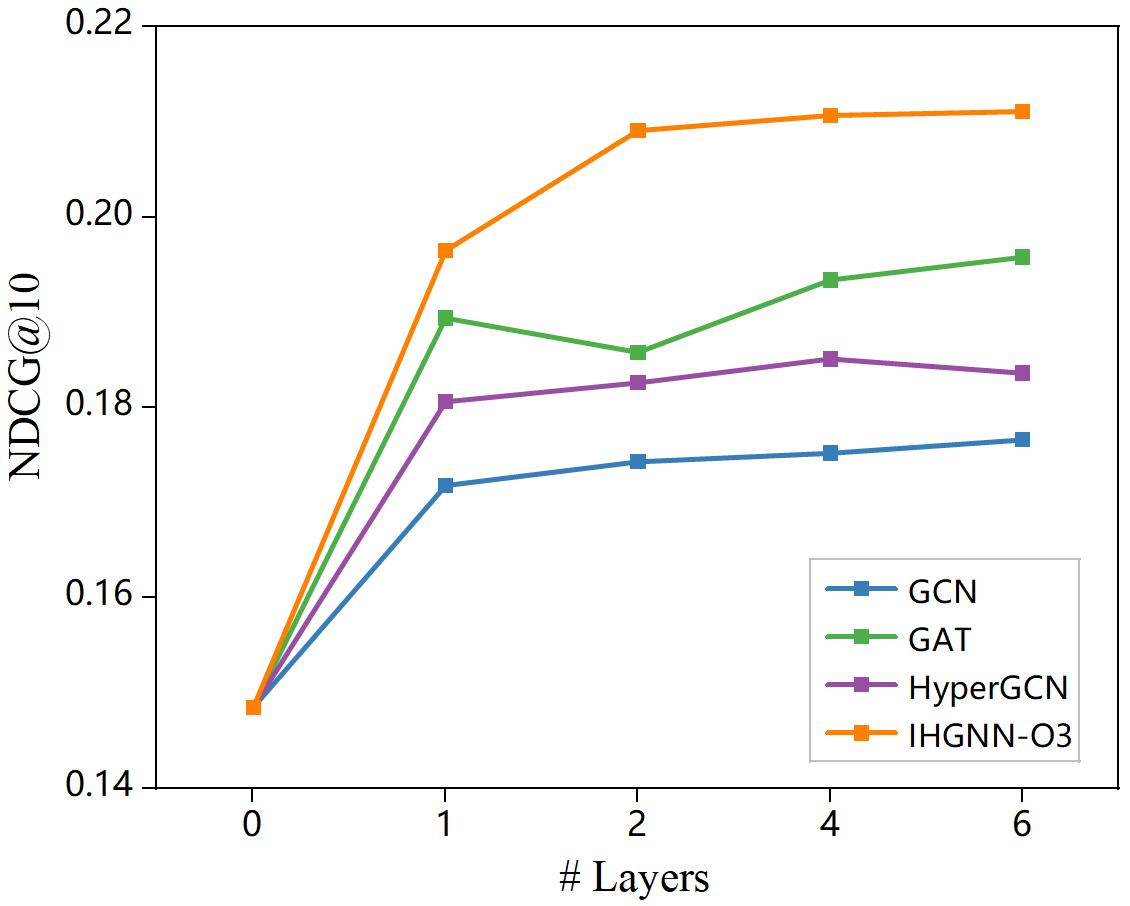}
  }
  \subfigure[Ali-1Core NDCG@10]{
    \label{fig_ablation_gnns_1C_NDCG}
    \includegraphics[width=0.235\linewidth]{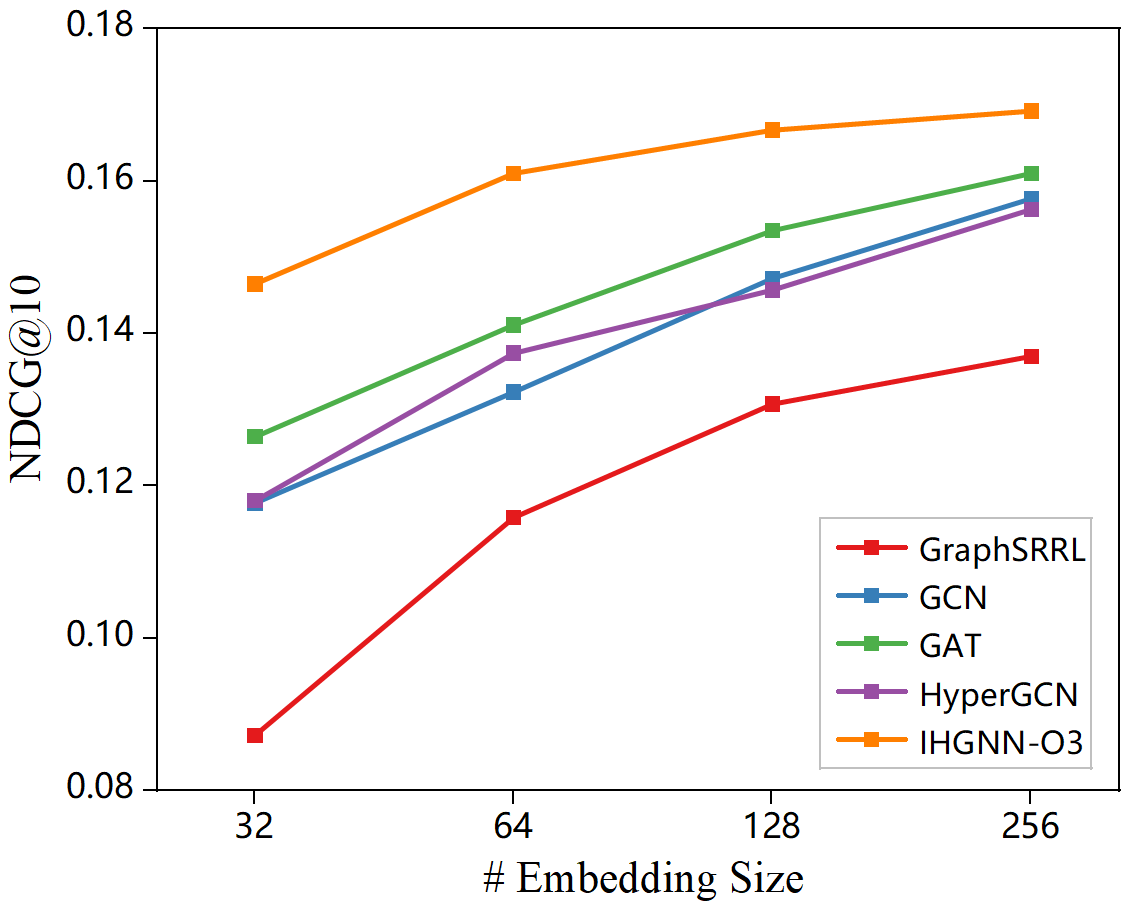}
  }
  \subfigure[Ali-5Core NDCG@10]{
    \label{fig_ablation_gnns_5C_NDCG}
    \includegraphics[width=0.235\linewidth]{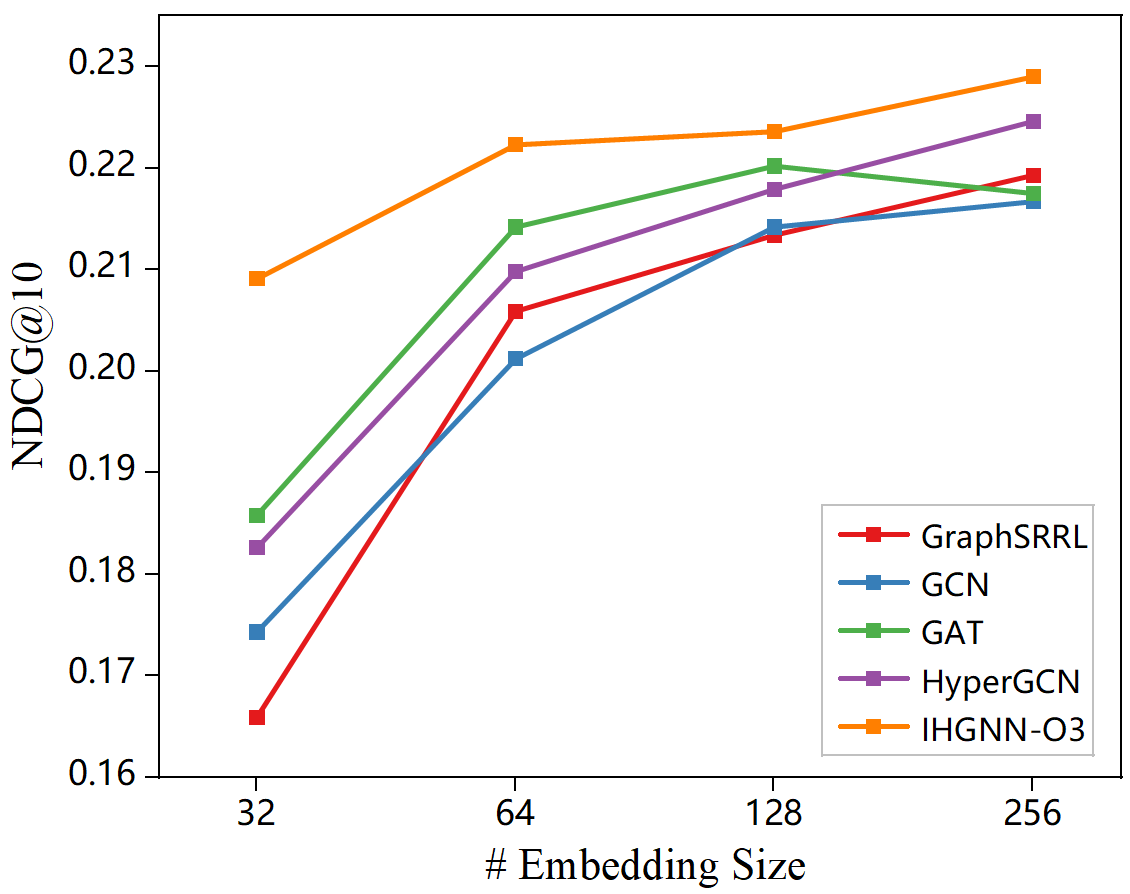}
  }
  \caption{Performance with different numbers of graph layers and different embedding sizes on Ali-1/5Core dataset.}
	\label{ablation_gnns}
\end{figure*}

\subsection{Performance Comparison (RQ1)}

Table \ref{ret_overall} presents the performance our IHGNN comparing with baselines. From the results, we have the following observations:


\textbf{Personalized vs. non-Personalized.} LSE has the worst performance. HEM is a personalized extension of LSE and has much better performance, which demonstrates that personalization is effective. This result is easy to understand because product search is a very personal behavior.


\textbf{Graph-based vs. not Graph-based.} Among all the baselines, methods based on graph usually have better performance. For example, on Ali-1Core, GCN performs 23.0\% better than methods without graph and GAT performs 32.1\% better in terms of NDCG@10. On other datasets, GCN and HyperGCN usually have good performance, and GAT always performs better than other baselines. This is because: 1) GNNs describe the relation among user, query and product very well, which is helpful for better personalization. 2) GNNs naturally model the 
interaction among users, queries, or products through multi-hop neighbors. HEM, ZAM and TEM does not exploit interactions in the user-query-product triples, and the meta-path-like method used by GraphSRRL does not directly utilize the interaction relation.


\textbf{IHGNN vs. Others.} Our proposed IHGNN significantly outperforms the baselines on all datasets. This is because:
1) Hypergraphs completely preserve the information in triple interaction data. Although HyperGCN does not outperform GCN at a significant level on Ali-1Core, the improvement from HyperGCN to IHGNN is much more than improvement from GCN to GAT. This indicates that the structure of hypergraph provides better ability to mine the personalization information in data.
2) We introduce weights in node aggregation. As different type of nodes may carry different levels of information, using weights would improve the performance. Similar results can be seen from the better performance of GAT over GCN.
3) Introduction of high-order feature interaction gives our model the ability to capture feature interactions that generally exist in the datasets.

\textbf{HyperGCN vs. GCN.} To our surprise, we can find HyperGCN performs quite close to GCN, even performs worse in some datasets. This result validates that although the hypergraph carries rich collabarative signal, we need to carefully design the embedding model to enjoy the merit of it. Blindly use existing methods may not bring much performance gain. Our IHGNN is specifically designed for PPS and achieves better performance than HyperGCN and GCN.

\begin{table}[t!]
  \centering
  \caption{
    Variants of IHGNN for ablations study.
    ``Node'' means node types in the graph, which contains options of user(u), query(q), and product(p).
    ``Weighted prop.'' points out if the model has mechanism of weighted propagation.
    ``Feature order'' refers to the maximum feature interaction order of the model.
  }
    \begin{tabular}{c|cccc}
    \toprule
          & Node & Weighted prop. & Feature order \\
    \midrule
    IHGNN-up & u, p   & $\times$     & 1 \\
    IHGNN-qp & q, p    & $\times$     & 1 \\
    HyperGCN & u, q, p  & $\times$    & 1 \\
    IHGNN-O1 & u, q, p  & $\surd$     & 1 \\
    IHGNN-O2 & u, q, p  & $\surd$     & 2 \\
    IHGNN-O3 & u, q, p  & $\surd$     & 3 \\
    \bottomrule
    \end{tabular}
  \label{tab:characteristics}
\end{table}


\textbf{Comparison across datasets.} As mentioned in section 4.1, real-world search data are usually very sparse. Therefore, we generate Ali-1Core to test models' performance on sparse data. From the table we can see that our IHGNN still yields a strong result. From Ali-5Core to Ali-1Core, baselines such as LSE and GraphSRRL decrease by 40.5\% and 42.2\% on average on HR@10 and NDCG@10. Baselines using GNN (GCN, GAT, HyperGCN) decrease by 30.8\% and 33.3\% on average. In contrast, our IHGNN only decrease by 28.4\% and 29.9\% on the two metrics. It shows that our model degrades less on sparser data. 

\subsection{Ablation Study (RQ2)}

In Table~\ref{tab:characteristics}, we listed different variants of IHGNN used for ablation study. The results of ablation study is shown in Table~ \ref{tab:ablation}.

\textbf{Contribution of weighted node aggregation.} The main difference between HyperGCN and IHGNN-O1 is whether to use weights in node aggregation. By comparing their results in Table \ref{tab:ablation}, we observe that the weighted propagation brings big performance promotion. This is because weighted aggregation allows our model to learn different importance of user, query and product on different datasets. It also helps to learn the importance of different features.

\textbf{Contribution of modeling high-order interaction.} By comparing the results of IHGNN-O1$\sim$3 in Table \ref{tab:ablation}, we can find considering 2- and 3-order feature indeed boosts PPS performance. This result is coincident with our intuition. The interactions among neighbors indeed provide useful signal to enhance the representation learning of the target node.

\textbf{Contribution of leveraging complete relations.} Section 4.4 has discussed the effect of using hypergraph comparing with collapsed graph. In this section, we further explore the effect of using complete relations. We construct a model that only consider user and product nodes: IHGNN-up, and another model that only considers query and product nodes: IHGNN-qp. 
By comparing IHGNN-up/qp with IHGNN-O1$\sim$3 in Table \ref{tab:ablation}, we see that modeling user-query-product triples using a hypergraph is much better than only modeling user-product or query-product tuples in product search scenario.

\begin{table}[t!]
  \centering
  \caption{Results of ablation study.}
    \begin{tabular}{c|cc|cc}
    \toprule
          & \multicolumn{2}{c|}{Ali-1Core} & \multicolumn{2}{c}{Ali-5Core} \\
          & HR@10 & NDCG@10 & HR@10 & NDCG@10 \\
    \midrule
    IHGNN-up & 0.1523  & 0.0997  & 0.2711  & 0.1893  \\
    IHGNN-qp & 0.1750  & 0.1171  & 0.2410  & 0.1637  \\
    HyperGCN & 0.1803  & 0.1180  & 0.2655  & 0.1826  \\
    IHGNN-O1 & 0.2038  & 0.1427  & 0.2812  & 0.2035  \\
    IHGNN-O2 & 0.2023  & 0.1420  & 0.2873  & \textbf{0.2094 } \\
    IHGNN-O3 & \textbf{0.2073 } & \textbf{0.1465 } & \textbf{0.2894 } & 0.2091  \\
    \bottomrule
    \end{tabular}
  \label{tab:ablation}
\end{table}

\subsection{Sensitivity Analysis (RQ3)}


\textbf{Effect of GNN layer count}. The number of GNN layers decides how many hops each node can visit in the process of message passing. In this experiment, we vary the number of GNN layers from 0 to 6 and test the performance of different models including GCN, GAT, HyperGCN and IHGNN-O3. The result is presented in Figure \ref{ablation_gnns} (a,b).
With the layer of the convolutions increasing, the performance will become better at the beginning. This result validates the effectiveness of leveraging (hyper)-graph aggregation in the recommender system. But when the layer surpasses a threshold, the performance becomes unaffected or even experiences some degradation with the further increase. Too deep layer may not bring additional collaborative signal, and even may bring some noise.


\textbf{Effect of embedding size}. We have also studied the effect of embedding size on models' performances on Ali-1Core and Ali-5Core data. We varied the embedding sizes from 32 to 256. The result is shown in Figure \ref{ablation_gnns} (c, d). The results show that increasing embedding size benefits IHGNN model. And our IHGNN model outperforms all baselines with different embedding sizes. We also observe that the superiority of IHGNN is more obvious on the sparser Ali-1Core dataset, which demonstrates that our model has better ability to capture feature information from sparse data.

\section{Related Work}
There are three lines of research directly related to our work: product search, graph based product search, and hypergraph learning.

\subsection{Product Search}

Early product search is based on structural data retrieved from relational databases, such as brands and categories \cite{lim2010multi,vandic2013facet}. 
\citeauthor{duan2013probabilistic}~\cite{duan2013probabilistic,duan2013supporting} proposed a probabilistic mixture model by analysing e-commerce search logs to effectively match query and title. 
But there are still semantic gaps between query words and product titles, which leads to the semantic mismatch problem. 
Some works~\cite{van2016learning, huang2013learning, shen2014latent} proposed to map query and product text into a hidden semantic space to learn the representations using deep neural networks.


Other than semantic match, personalization is playing a important role in improving user experience and increasing the retailer revenues in product search.
\citeauthor{hem}~\cite{hem} and \citeauthor{ge2018personalizing}~\cite{ge2018personalizing} both proposed a hierarchical embedding model to jointly learn representations of users, queries, and products in a hidden semantic space. 
A zero attention model (ZAM)~\cite{zam} is then proposed that automatically determines the level of personalization for a specific user-query pair.
\citeauthor{tem}~\cite{tem, rtm} proposed a transformer-based embedding model (TEM) and a review-based transformer model (RTM) to dynamically control the influence of personalization.
\citeauthor{zhou2020encoding}~\cite{zhou2020encoding} proposed a context-aware long-short term preference model to enhance the representation of the current query.

Some researchers focus on different scenarios in product search, such as streaming~\cite{xiao2019dynamic} and conversational systems~\cite{zhang2018towards,bi2019conversational}. 
Some works leverages different information in product search to help ranking. 
\citeauthor{long2012enhancing}~\cite{long2012enhancing} used popularity combined with relevance to improve product ranking. 
\citeauthor{guo2018multi}~\cite{guo2018multi} utilized multi-modal data in product search. 
\citeauthor{wu2018turning}~\cite{wu2018turning} utilized click and purchase data to jointly optimize search results.

\subsection{Graph Based Product Search}

Graph embedding method has been proven effective in information retrieval task \cite{hamilton2017inductive}. 
\citeauthor{drem}~\cite{drem} proposed to construct explainable PPS model through a product-category-product knowledge graph. Similarly, \citeauthor{niu2020dual}~\cite{niu2020dual} construct a dual heterogeneous graph attention network for the query-item graph in e-commerce search.
\citeauthor{zhang2019neural}~\cite{zhang2019neural} used graph-based features to enhance learning-to-rank frameworks.
\citeauthor{ren2014heterogeneous}~\cite{ren2014heterogeneous} used heterogeneous click graph to learn generic search intents. 

We argue that existing graph-based methods only include two types of entities ((query,product) or (user,product)) rather than all of them (user,product,query) in the constructed graph, failing to modeling the ternary relation among them.  Our experiments also validate that modeling ternary relations is quite useful.
To the best of our knowledge, only one work \citeauthor{srrl}~\cite{srrl} considered all the entities in the graph and exploited there types of structural relationship in user-query-product interactions. But such three manually-designed patterns are far from sufficient.

\subsection{Hypergraph Learning}
Relations of three or more entities cannot be represented by a traditional graph.
Hypergraph naturally models high-order relations. 
\citeauthor{hgraph}~\cite{hgraph} first introduced hypergraph learning into transductive classification task, generalizing the methodology of spectral clustering to hypergraphs, which originally operates on undirected graphs. 
\citeauthor{hgcn}~\cite{hgcn} presented hypergraph neural networks (HGNNs) to consider the high-order data structure to learn representation better with multi-modal data. 
\citeauthor{mao2019multiobjective}~\cite{mao2019multiobjective} applied hypergraph ranking to multi-objective recommendation task by establishing a user-product-attribute-context data model. 
\citeauthor{wang2020next}~\cite{wang2020next} developed a next-item recommendation framework empowered by sequential hypergraphs to infer the dynamic user preferences with sequential user interactions. 

However, to our best knowledge, few works used hypergraph in personalized product search.
In this paper, we applied hypergraph neural networks to PPS task and improved our model's performance by optimizing model structure and exploiting high-order feature interactions. Also, we remark these methods may not be suitable to be directly applied in PPS. Our IHGNN is designed specifically for PPS with modeling weighs in node aggregation and modeling high-order feature interactions. Although there are some recent work \cite{zhu2020bilinear, feng2021cross} modeling high-order feature interaction on GNN, the feature interaction on hypergraph has not been explored.
\section{Conclusions and Future Work}
In this article, we proposed a new hypergraph-based method IHGNN to better model the crucial \textit{collaborative signal} among users, queries and products for personalized product search. 
Also, we developed an improved hypergraph neural network and explicitly introduced the neighbor feature interactions.
Extensive experiments have been conducted to validate the superiority of IHGNN over baselines.
We also analyze the contribution of each component in our model.
For future work, we consider using attention mechanisms in both hypergraph propagation phases and high-order feature calculation. Also, it would be interesting to explore disentangling representations in PPS, as users may have diverse preference and queries may have diverse semantics. 

\section{Acknowledgements}
This work is supported by the National Natural Science Foundation of China (62102382,U19A2079), the USTC Research Funds of the Double First-Class Initiative (WK2100000019) and the Alibaba Innovative Research project (ATT50DHZ420003).

\bibliographystyle{ACM-Reference-Format}
\balance 
\bibliography{sigproc.bib}

\appendix
\end{document}